\documentstyle[preprint,aps]{revtex}
\begin{document}
\draft
%\twocolumn[\hsize\textwidth\columnwidth\hsize\csname %
%@twocolumnfalse\endcsname

\title{Alternating chain with Hubbard-type interactions:
renormalization group analysis}
\author{F.D. Buzatu$^{a}$, G. Jackeli$^{b,c}$}
\address{$^{a}$ Department of Theoretical Physics,
 Institute for Physics and Nuclear Engineering \\
 P.O.Box Mg-6, M\u{a}gurele, Bucharest, R-76900, Romania\\
$^{b}$Joint Institute for Nuclear Research, Dubna, Moscow region,
 141980, Russia \\
$^{c}$ Department of Solid State Physics,
Tbilisi State University, Tbilisi, Georgia.}
\maketitle
\begin{abstract}\widetext
The canonical transformation diagonalizing the one-particle tight binding
Hamiltonian for an alternating chain with two non-equivalent sites per unit
cell has been used to introduce the Hubbard-like interactions (on-site,
inter-site, bond-site and intra-bond) in the corresponding two-band model.
The considerations have been restricted to the particular case, suggested
by the undistorted 3/4-filled $CuO_{3}$ chain, where the alternating
structure comes only from non-equal on-site atomic energies and
the gap between the two bands is sufficiently large.
The renormalization group method has been then applied to the
upper band of this model and the corresponding phase diagram has been
analyzed in terms of some renormalized (density dependent) Hubbard-type
couplings for arbitrary filling of the upper band.
The particular case of $CuO_{3}$ chain is also discussed and
the corresponding phase diagram has been drawn in terms of the original
(and realistic) coupling constants.
\end{abstract}

%]
\narrowtext

\section{Introduction}

A large amount of work has been devoted to the study of
alternating chains not only due to their interesting
structure, but also in connection with a better understanding of the
high-$T_{c}$ superconductivity mechanism ~\cite{DME}.
The same phenomenon renewed the interest in the Hubbard model and,
quite recently, in its one-dimensional extensions ~\cite{ext}.
However, the particularities induced by the Hubbard-like couplings in
alternating chains are less well known:
the charge gap that opens in a one-dimensional dimerized Hubbard model
has been estimated in various limits using analytical results and
exact diagonalizations of small clusters ~\cite{PM};
a model where the dimerization is induced via alternating on-site
repulsions has been studied within a boson representation theory ~\cite{JKM}
and also by renormalization group (RG) technique ~\cite{JJ}.
The aim of this paper is to investigate, using the RG method,
the effect of the Hubbard-type interactions on the
ground-state properties of a chain with alternating on-site atomic energies.

The paper is structured as follows. The one-particle Hamiltonian
in the tight binding approximation corresponding to an alternating chain
with two nonequivalent sites per unit cell can be diagonalized by a
canonical transformation; one gets  a two band model. The Hubbard-type
interactions, i.e. interactions which in the
site representation couple only electrons belonging to the
nearest neighbor sites, give rise to both intra- and
inter-band couplings; however, if the gap between the two bands
is sufficiently large and the system is more than half-filled,
as for the $CuO_{3}$ chain occurring in high-$T_c$ superconductors,
the last ones can be neglected in describing the low energy physics.
The obtained expressions of the Hubbard-type interactions (upper band),
in the particular case of alternating on-site energies and equal hopping
amplitudes, close Sec. II. The standard RG analysis (second order) ~\cite{S},
briefly reviewed in Sec III, is done in terms of the
$g$-constants describing the elementary processes of forward, backward and
umklapp scatterings: their expressions are obtained by evaluating the
Hubbard-type interactions (upper band) at the Fermi points.
Using the scaling to the exact soluble models
Tomonaga-Luttinger (TL)~\cite{TL} and Luther-Emery (LE) ~\cite{LE},
we can predict the low energy physics of our system; the ground-state
phase diagrams in terms of the model parameters and at arbitrary band
filling are presented in Sec. IV. The relevance of our analysis to
the case of the undistorted 3/4-filled $CuO_{3}$ chain and the corresponding
phase diagram in terms of the original coupling constants (neglecting
the intra-bond coupling which usually is very small) are discussed in
Sec. 5. The last section summarizes the main results of this work.

\section{The two-band model}

Let us consider the alternating chain from Fig. 1, with
two non-equivalent sites per unit cell ($N$ cells, periodic boundary
conditions). Within the {\em tight binding approximation},
the second quantized form of the one-particle Hamiltonian in the
{\em site representation} of the atomic orbitals reads~\cite{BP}
\begin{eqnarray}
{\cal H}_{0} & = & \epsilon
\sum\limits_{j}^{}\left( a^{\dag}_{j}a^{}_{j}-b^{\dag}_{j}b^{}_{j}\right)
\nonumber \\
& & \mbox{} - \sum\limits_{j}^{}\left[ \left( ta^{\dag}_{j}+
\overline{t}a^{\dag}_{j+1}
\right) b^{}_{j}+H.c.\right]
\end{eqnarray}
where $j=\overline{1,N}$ is the cell index, and the
$a(b)$-operator corresponds to the annihilation of one electron on
a site $A(B)$; we ignore for the moment
the spin variable $\sigma =\uparrow ,\downarrow $ of the electron
(it can also be considered as included in the cell index).
The constants $\epsilon $ and $t$ ($\overline{t}$) are positive and
denote respectively the site energies
and the parameter describing the hopping between $A$ and $B$ sites
belonging to the same cell (nearest neighbor cells); all of them can
be expressed in terms of the atomic quantities ~\cite{BP}.
From the site representation we can pass to the {\em momentum representation}
by a usual Fourier transform; Eq. (1) becomes then
\begin{eqnarray}
{\cal H}_{0} & = & \epsilon
\sum\limits_{k}^{}\left( a^{\dag}_{k}a^{}_{k}-b^{\dag}_{k}b^{}_{k}\right)
\nonumber \\
& & \mbox{} - \sum\limits_{k}^{}\left[ \left( t+\overline{t} e^{-iak}\right)
a^{\dag}_{k}b^{}_{k}+H.c.\right]
\end{eqnarray}
where $k\in (-\pi /a,\pi /a]$ with $a$ being the lattice constant.

By mixing the $a$- and $b$-operators, the Hamiltonian ${\cal H}_{0}$
has not a diagonal form; it can be diagonalized
by the canonical transformation~\cite{BP}
\begin{equation} \left\{
 \begin{array}{l}
 a^{}_{k}=A(k)e^{-i\phi(k)}c^{}_{1,k}+B(k)c^{}_{2,k} \\
			  \\
 b^{}_{k}=B(k)c^{}_{1,k}-A(k)e^{i\phi(k)}c^{}_{2,k}
 \end{array} \right.
\end{equation}
with
\begin{equation}
\left\{ \begin{array}{l}
	 {\displaystyle A(k)=\frac{1}{\sqrt{2}}\left[ 1-
  \frac{\epsilon}{\varepsilon(k)}
  \right]^{\frac{1}{2}}}  \\
	 \\
	 {\displaystyle B(k)=\frac{1}{\sqrt{2}}\left[ 1+
  \frac{\epsilon}{\varepsilon(k)}
  \right]^{\frac{1}{2}}}  \\
	 \\
	  {\displaystyle \tan[\phi(k)]=\frac{\overline{t}\sin(ak)}{t+
				    \overline{t}\cos(ak)}}
	 \end{array}
\right.
\end{equation}
$\varepsilon(k)$ in Eq. (4) has the form
\begin{equation}
\varepsilon(k)= \sqrt{\Delta^{^{2}}+4t\overline{t}\cos^{2}(ak/2)}
\end{equation}
where
\begin{equation}
\Delta =\sqrt{\epsilon^{2}+\left(t-\overline{t}\right)^{2}}
\end{equation}
In terms of the $c$-operators,
the one-particle Hamiltonian (2) reads as
\begin{equation}
{\cal H}_{0} = \sum\limits_{k,\alpha}^{} (-1)^{\alpha} \varepsilon(k)
c^{\dag}_{k,\alpha}c^{}_{k,\alpha} \ , \  \alpha =1,2
\end{equation}
and defines the kinetic part of the two-band model:
the dispersion law in the upper (lower) band is given by plus (minus)
$\varepsilon(k)$ with a gap between the two bands equal to
$2\Delta$; in the limit $\Delta \rightarrow 0$ , one recovers the usual
dispersion law $-2t\cos(ak)$ for an ideal (non-alternating) structure.

By analogy with the usual case, the Hubbard-type interactions
between the electrons in an alternating chain are introduced
in the site representation as follows:

(i) {\em on-site}
\begin{equation}
{\cal H}_{1}=\frac{1}{2}\sum\limits_{j,\sigma}^{}\left(
U_{A}n^{A}_{j,\sigma}n^{A}_{j,-\sigma}+
U_{B}n^{B}_{j,\sigma}n^{B}_{j,-\sigma}\right)
\end{equation}
where $n^{A(B)}_{j,\sigma}=
a^{\dag}_{j,\sigma}a^{}_{j,\sigma}
(b^{\dag}_{j,\sigma}b^{}_{j,\sigma})$;

(ii) {\em inter-site}
\begin{equation}
{\cal H}_{2}=\sum\limits_{j,\sigma ,\sigma^{\prime}}^{}
n^{A}_{j,\sigma}\left(
Vn^{B}_{j,\sigma^{\prime}}+\overline{V}n^{B}_{j-1,\sigma^{\prime}}\right)
\end{equation}

(iii) {\em bond-site}
\begin{eqnarray}
{\cal H}_{3}&=&\sum\limits_{j,\sigma }^{}
\left[ a^{\dag}_{j,\sigma}\left(
X_{A}b^{}_{j,\sigma}+\overline{X}_{A}b^{}_{j-1,\sigma}\right) + H.c.
\right] n^{A}_{j,-\sigma}                \nonumber \\
&+ & \sum\limits_{j,\sigma}^{}
\left[ b^{\dag}_{j,\sigma}\left(
X_{B}a^{}_{j,\sigma}+\overline{X}_{B}a^{}_{j+1,\sigma}\right) + H.c.
\right] n^{B}_{j,-\sigma}
\end{eqnarray}

(iv) {\em exchange-hopping}
\begin{eqnarray}
{\cal H}_{4} & = &
-\frac{1}{2}\sum\limits_{j,\sigma ,\sigma^{\prime}}^{}
\left[ a^{\dag}_{j,\sigma}\left(
Wb^{\dag}_{j,\sigma^{\prime}}b^{}_{j,\sigma} \right. \right. \nonumber \\
  & & \mbox{} + \left. \left.
\overline{W}b^{\dag}_{j-1,\sigma^{\prime}}b^{}_{j-1,\sigma}\right)
a^{}_{j,\sigma^{\prime}} + H.c. \right]
\end{eqnarray}

(v) {\em pair-hopping}
\begin{eqnarray}
{\cal H}_{5} & = &
\frac{1}{2}\sum\limits_{j,\sigma}^{}
\left[ a^{\dag}_{j,\sigma}a^{\dag}_{j,-\sigma}\left(
Wb^{}_{j,-\sigma}b^{}_{j,\sigma} \right. \right. \nonumber \\
 & & \mbox{} + \left. \left.
\overline{W}b^{}_{j-1,-\sigma}b^{}_{j-1,\sigma}\right) + H.c. \right]
\end{eqnarray}

The expressions of the Hubbard-type interactions in the corresponding
two-band model are obtained by passing in Eqs. (8)-(12) to the momentum
representation and replacing after that the $a^{}_{k}$- and
$b^{}_{k}$-operators by the $c^{}_{\alpha,k}$-operators, according to Eq. (3).
Below we shall restrict our considerations only to the
particular case of alternating on-site
energies ($\epsilon \neq 0$) and equal hopping amplitudes
($t=\overline{t}$); in this case ($a_{1}=a_{2}=a/2$)
\begin{equation}
\phi =\frac{ak}{2} \ \ ,\ \ \Delta =\epsilon
\end{equation}
and there  will be no distinction between intra-cell interaction constants
(without overline) and the inter-cell ones (with overline).
Any interaction in terms of the $a$- and $b$-operators gives rise to both
intra- and inter-band couplings. Nevertheless, if the gap between
the two bands is large enough (comparable to the bandwidth and larger than
all electron couplings) the considerations can be restricted
to the partial filled band; assuming this fact together with a concentration
greater than one electron per site, we shall consider further only
processes from the upper band (the realistic case of the
$CuO_{3}$ chain will be discussed in Sec. V).
Our (one-band) model Hamiltonian reads thus as (the index $\alpha=2$ will
be omitted in $c$-operators)
\begin{eqnarray}
{\cal H} & = & \sum\limits_{k,\sigma}^{} \varepsilon(k)
c^{\dag}_{k,\sigma}c^{}_{k,\sigma}
 + \frac{1}{2N} \sum\limits_{\beta =1}^{5}
 \sum\limits_{k_{1-4};\sigma ,\sigma^{\prime}}^{}
 \delta^{}_{k_{1}+k_{2},k_{3}+k_{4}} \nonumber  \\
  & & \mbox{} \times V_{\beta}(k_{1},..,k_{4}; \sigma ,\sigma^{\prime})
 c^{\dag }_{k_{1},\sigma }c^{\dag }_{k_{2},\sigma^{\prime} }
 c^{}_{k_{4},\sigma^{\prime} }c^{}_{k_{3},\sigma }
\end{eqnarray}
where $\varepsilon(k)$ is given by Eq. (5);
$V_{\beta}$ quantities ($\beta=\overline{1,5}$) correspond, respectively, to
the five types of the Hubbard interactions defined by
Eqs. (8)-(12), and they have the following expressions:
\begin{equation}
V_{1}=\left[ U_{A} \prod\limits_{i=1}^{4}B(k_{i}) \pm
    (A\leftrightarrow B) \right] \delta_{\sigma ,-\sigma^{\prime}}
\end{equation}
\begin{eqnarray}
V_{2} & = & 2V\cos[a(k_{1}-k_{3})/2]  \nonumber  \\
      & & \mbox{} \times \left[ A(k_{1})B(k_{2})A(k_{3})B(k_{4})
 \pm (A\leftrightarrow B) \right]
\end{eqnarray}
\begin{eqnarray}
V_{3} & = & -4\left\{ X_{A} \left[
\cos(ak_{1}/2) A(k_{1})B(k_{3}) + \cos(ak_{3}/2)
  \right. \right. \nonumber  \\
  & \times & \left. \left.
 B(k_{1})A(k_{3}) \right] B(k_{2})B(k_{4}) \pm
    (A\leftrightarrow B) \right\} \delta_{\sigma ,-\sigma^{\prime}}
\end{eqnarray}
\begin{eqnarray}
V_{4} & = & 2W\cos[a(k_{1}-k_{4})/2]  \nonumber  \\
 & & \mbox{} \times \left[ A(k_{1})B(k_{2})B(k_{3})A(k_{4}) \pm
  (A \leftrightarrow B) \right]
\end{eqnarray}
\begin{eqnarray}
V_{5} & = & 2W\cos[a(k_{1}+k_{2})/2]  \nonumber \\
  & & \times \left[ A(k_{1})A(k_{2})B(k_{3})B(k_{4}) \pm
  (A\leftrightarrow B) \right]
  \delta_{\sigma ,-\sigma^{\prime}}
\end{eqnarray}
The $\delta $-function in Eq. (14) assures the conservation of
the total momentum up to a reciprocal lattice vector, i.e.
 $k_{1}+k_{2}=k_{3}+k_{4}+Q$ with $Q=0$ or $Q=\pm 2\pi /a$.
In Eqs. (15)-(19), the upper (lower) sign corresponds to the normal
(umklapp) scattering, i.e. to $Q=0$ $(\pm 2\pi /a)$; this fact,
characteristic of an alternating structure ~\cite{PM}, comes from
the phase factor $\phi$ of the canonical transformation (3).
Let us also note that for the lower band ($\alpha =1$),
$V_{2}$, $V_{4}$ and $V_{5}$ have the same expressions while
$V_{1} \rightarrow V_{1}\left[ U_{A} \leftrightarrow U_{B} \right] $ and
$V_{3} \rightarrow -V_{3}\left[ X_{A} \leftrightarrow X_{B} \right] $.

\section{Renormalization group analysis}

The low energy physics of our model (14) can be described, within the
RG method ~\cite{S}, by assuming: (i) a linear dispersion
law
\begin{equation}
\varepsilon(k)\simeq v_{F}(k_{F}-|k|) \ , \ v_{F} >0
\end{equation}
together with the existence of a momentum cut-off $k_{0}$ restricting
all possible states of the electrons to those around the Fermi points
(or equivalently, a bandwidth cut-off $E_{0}=2v_{F}k_{0}$); (ii) all
interaction processes can be classified into four different types
%(see Fig. 3)
with the coupling constants  $g^{}_{i}$ ($i=\overline{1,4}$)
obtained by evaluating the bare potentials at the corresponding values
$+$ or $-k_{F}$ of the momenta.

With the usual notations ~\cite{S},
the expressions of the $g$-constants corresponding to the model (14)
and also their form for a non-alternating chain are given in Table I,
where we have introduced the renormalized Hubbard constants
\begin{equation} \left\{
   \begin{array}{l}
{\cal U}=U_{A}B^{4}(k_{F}) + U_{B}A^{4}(k_{F}) \\
 \\
{\cal V}=2VA^{2}(k_{F})B^{2}(k_{F})  \\
  \\
{\cal X}=X_{A}A(k_{F})B^{3}(k_{F}) +
  X_{B}A^{3}(k_{F})B(k_{F}) \\
 \\
{\cal W}=2WA^{2}(k_{F})B^{2}(k_{F})
   \end{array}   \right.
\end{equation}
Unlike for a non-alternating chain, they now depend on the
band filling $n$ [$ak_{F}=\pi (1-n)$, $0\leq n\leq 1$].
As can be remarked from Table I, there is an almost complete analogy
between the case of an $n$-filled band coming from an alternating chain and
a ($n$+1)/2-filled band of a non-alternating one; the differences
consist in the renormalization of the Hubbard constants and, obviously,
in the umklapp process.

All physical results predicted by the RG method
(second order) can be discussed in terms of four independent coupling
constants: $g^{}_{1\perp}$, $g^{}_{3}$,
\begin{equation}
g^{}_{\sigma}=g^{}_{1{\scriptscriptstyle \parallel}}-
g^{}_{2{\scriptscriptstyle \parallel}}+g^{}_{2\perp}
\end{equation}
and
\begin{equation}
g^{}_{\rho}=g^{}_{1{\scriptscriptstyle \parallel}}-
g^{}_{2{\scriptscriptstyle \parallel}}-g^{}_{2\perp}
\end{equation}
(the main effect of $g^{}_{4}$ can be included in a renormalized Fermi
velocity). The coupling constants
$g^{}_{\sigma}$ and $g^{}_{1\perp}$ describe the spin sector;
$g^{}_{\rho}$ and $g^{}_{3}$, the charge sector.
The RG equations (in units of $\pi v_{F}$) read as~\cite{S}
\begin{equation} \left\{
  \begin{array}{l}
{\displaystyle \frac{d g^{}_{\sigma}}{d x}}(x)=g_{1\perp}^{2}(x)
[1+\frac{1}{2}g^{}_{\sigma}(x)] \\
\\
{\displaystyle \frac{d g^{}_{1\perp}}{d x}}(x)=
g^{}_{\sigma}(x)g^{}_{1\perp}(x)[1+
\frac{1}{4}g^{}_{\sigma}(x)]+\frac{1}{4}g_{1\perp}^{3}(x)
  \end{array} \right.
\label{L1}
\end{equation}
\begin{equation}  \left\{
  \begin{array}{l}
{\displaystyle \frac{d g^{}_{\rho}}{d x}}(x)=g_{3}^{2}(x)
[1+\frac{1}{2}g^{}_{\rho}(x)] \\
\\
{\displaystyle \frac{d g^{}_{3}}{d x}}(x)=g^{}_{\rho}(x)g^{}_{3}(x)[1+
\frac{1}{4}g^{}_{\rho}(x)]+\frac{1}{4}g_{3}^{3}(x)
  \end{array} \right.
\label{L2}
\end{equation}
and reflects the charge-spin separation;
$x=\ln(E/E_{0})\in (-\infty ,0]$, with $E$ a smaller cut-off
than the original one $E_{0}$.
By solving Eqs. (\ref{L1}) and (\ref{L2}),
a set of equivalent problems, related by RG transformations,
can be found; the low energy physics is essentially the same for all models
with the $g$-constants along a certain $g(x)$ solution.
The corresponding flow diagrams indicate two distinct regimes:
(i) for $g^{}_{\sigma}\geq |g^{}_{1\perp}|$ ($g^{}_{\rho}\geq |g^{}_{3}|$)
the spin (charge) part of the system scales to the exact soluble
TL model~\cite{TL}
which is also the fixed point (this is the weak coupling regime);
(ii) for $g^{}_{\sigma}<|g^{}_{1\perp}|$ ($g^{}_{\rho}< |g^{}_{3}|$) the spin
(charge) part of the system scales to one of the two
strong coupling fixed points, but before this it crosses the LE
line and thus its behavior can be inferred from the exact solution~\cite{LE}.

Let us note that for the Hubbard-type interactions, due to the
$SU(2)$-spin symmetry of the Hamiltonian (14),
$g^{}_{\sigma}$ is always equal to $g^{}_{1\perp}$; consequently,
we get four distinct regions
\begin{equation} \begin{array}{cc}
I: \left\{  \begin{array}{l}
	     g^{}_{\sigma}\ge 0  \\
	     g^{}_{\rho}\ge |g^{}_{3}|
	    \end{array}
   \right.  &
II: \left\{  \begin{array}{l}
	     g^{}_{\sigma}\ge 0  \\
	     g^{}_{\rho}< |g^{}_{3}|
	    \end{array}
   \right.   \\
    &   \\
III: \left\{  \begin{array}{l}
	     g^{}_{\sigma}< 0  \\
	     g^{}_{\rho}< |g^{}_{3}|
	    \end{array}
   \right.  &
IV: \left\{  \begin{array}{l}
	     g^{}_{\sigma}< 0  \\
	     g^{}_{\rho}\ge |g^{}_{3}|
	    \end{array}
   \right.
 \end{array}
\end{equation}
where, in our case,
\begin{eqnarray}
g^{}_{\sigma} & = & {\cal U}-2{\cal V}\cos(\pi n)-8{\cal X}\sin(\pi n/2)+
4{\cal W}  \\
 & & \nonumber \\
g^{}_{\rho} & = & -{\cal U}-2{\cal V}[2+\cos(\pi n)]+
 8{\cal X}\sin(\pi n/2) \nonumber \\
  & & +4{\cal W}\cos(\pi n)
\end{eqnarray}
and $g^{}_{3}$ can be read from Table I.
In terms of the model parameters, the four regions defined by Eq. (26)
can be conveniently
described in the (${\cal W},{\cal V}$)-plane as follows:
\begin{equation} \begin{array}{cc}
I: \left\{  \begin{array}{l}
	     {\cal W}\ge F  \\
	     {\cal V}\le G
	    \end{array}
   \right.  &
II: \left\{  \begin{array}{l}
	     {\cal W}\ge F  \\
	     {\cal V}> G
	    \end{array}
   \right.   \\
    &   \\
III: \left\{  \begin{array}{l}
	     {\cal W}< F  \\
	     {\cal V}> G
	    \end{array}
   \right.  &
IV: \left\{  \begin{array}{l}
	     {\cal W}< F  \\
	     {\cal V}\le G
	    \end{array}
   \right.
 \end{array}
\end{equation}
where
\begin{equation}
F=\left[ -{\cal U}+2{\cal V}\cos(\pi n)
+8{\cal X}\sin(\pi n/2) \right] /4
\end{equation}
does not depend on ${\cal W}$ and
\begin{equation}
G=\left\{
 \begin{array}{ll}
{\displaystyle
 \frac{8{\cal X}\sin(\pi n/2)
+4{\cal W}\cos(\pi n)-{\cal U}}{2[2+\cos(\pi n)]}} & n\neq \frac{1}{2} \\
 & \\
 -\frac{1}{2}Max \left\{ Y_{1},Y_{2} \right\} & n=\frac{1}{2}
   \end{array} \right.
\end{equation}
is independent of ${\cal V}$ where
\begin{equation}
Y_{1}=U_{A}B^{4}-4\sqrt{2}X_{A}AB^{3} \ \ , \ \
Y_{2}=Y_{1}[A\leftrightarrow B]
\end{equation}
$A$ and $B$ in Eq. (32) stand respectively for $A(k_{F})$ and
$B(k_{F})$ [see Eq. (4)] evaluated at half-filling.
The scaling of our system to one of the two exactly soluble models,
TL or LE, in each of the four regions, at any density and for both
charge and spin sector is summarized in Table II.

\section{Phase diagrams}

Based on the exact results obtained for the TL and LE models and using the
scaling arguments, we can now describe the low energy physics of our
system. Anytime the system scales to the LE model, there is a gap in the
corresponding charge or spin sector; in the TL case, the spectrum is
gapless. Following the S\'{o}lyom's analysis ~\cite{S}, we can predict
the most preferred type of instability occurring in the ground-state
of the system, corresponding to the most divergent correlation function:
charge density wave ($CDW$), spin density wave ($SDW$),
singlet superconductivity ($SS$) or triplet superconductivity ($TS$).
Before discussing the results, let us remark that the on-site
and bond-site interaction constants have always opposite effects; they occur
only in the combination
\begin{equation}
Y={\cal U}-8{\cal X}\sin(\pi n/2)
\end{equation}
or, in the half-filled case, through $Y_{1}$ or $Y_{2}$ with
a similar structure. Consequently, a bond-site repulsion acts (in the
upper band) as an effective attraction and its effect is enhanced by
increasing the electron density $n$, a fact already used in the
hole superconductivity mechanism ~\cite{Hi}. The value of $Y$
(or $Y_{1,2}$) fixes the position, in the (${\cal W},{\cal V}$)-plane,
of the intersection between the ``spin-line" ($g^{}_{\sigma}=0$) and the
``charge-line" ($g^{}_{\rho}=|g_{3}|$).

Let us first consider the half-filled case ($n=1/2$). The obtained
phase diagram has the structure presented in Fig. 3. The ``spin-line"
is vertical and the ``charge-line" is horizontal;
both of them delimit not only different phases, but also the
strong coupling regime from the weak coupling one in each
corresponding (charge or spin) sector .
The critical value ${\cal V}_{2}=-Max\left\{ Y_{1},Y_{2}\right\}/2$
of the inter-site
interaction constant separates a dominant superconductor region
(${\cal V}<{\cal V}_{2}$) from a density fluctuation one
(${\cal V}>{\cal V}_{2}$). Analogously, by decreasing the intra-bond
interaction constant ${\cal W}$ below the critical value
${\cal W}_{1}=-Y/4$, the triplet states (spin-density or superconductor)
disappear.
Depending on the values of $U_{A,B}$ and $X_{A,B}$, the origin
${\cal W}={\cal V}=0$ can be found in principle in any of the four regions,
but the most probable case for a real system is that considered in the
picture ($Y>0$, $Y_{1,2}>0$).

In the non-half-filled case, $g_{3}$ is zero and consequently $g_{\rho}$
does not renormalize; thus
the charge sector is always in the weak coupling regime and
the ``charge-line" now separates only different phases. Graphically, two
things happen: (i) the intersection
between the ``charge-line" and the ``spin-line" moves in such a way
that the origin ${\cal W}={\cal V}=0$ lies inside either  the region II
(for $Y>0$) or the region IV (for $Y<0$); (ii) by increasing the
density $n$, the ``spin-line" rotates anti-clockwise while the
``charge-line" rotates clockwise (by decreasing $n$, the effect is reversed).
Consequently, on starting from the half-filled case and
increasing the electron concentration
the regions I and III become smaller while the
regions II and IV grow, as it is shown in Fig. 4 for $Y>0$; the intersection
of the ``spin-line" with the coordinate axes are given respectively by
${\cal W}_{1}=-Y/4$ and ${\cal V}_{1}=Y/[2\cos(\pi n)]$
while in the ``charge-line" case by
${\cal W}_{2}=Y/[4\cos(\pi n)]$ and ${\cal V}_{2}=-Y/[4+2\cos(\pi n)]$.
The effect of the model parameters on the ground-state phase diagram
can easily be inferred from the above discussion.

%%%%%%%%%%%%%%%%%%%%%%%%%%%%%%%%%%%%%%%%%%%%%%%%%%%%%%%%%%%%%%%%%%%%%%%%%%%%%

\section{The case of $CuO_{3}$ chain}

The use of the RG method implicitly assumes a weak coupling regime;
the scaling equations are obtained in the second order of the perturbation
theory and consequently the coupling constants have to be small
in comparison with the bandwidth.
For the 3/4-filled undistorted $CuO_{3}$ chain we have
$t=\overline{t}=1.4$ eV with a gap
$2\Delta = 2\varepsilon=1.23$ eV~\cite{DMLK} and from Eq. (5) it follows a
bandwidth around 2.25 eV (and thus a quite large gap). On the other hand,
$U_{A}\approx $3 - 4 eV, $U_{B}\approx $1 - 3 eV ~\cite{DMLK},
i.e. we are in an intermediate coupling regime; in this respect,
our estimations for the $CuO_{3}$ chain presented below could provide
only a qualitative description of the system. The restriction to the
partial filled (upper) band, justified also in the weak coupling limit,
supports the same conclusion.

Excepting the on-site interaction, the values of the other coupling
constants for the $CuO_{3}$ chain are less known.
For a non-alternating chain, using a Kronig-Penny model for the ionic
potential and a screened Coulomb inter-electronic potential, Campbell
et al.~\cite{PG,CGL} analyzed the relative magnitude of the Hubbard-type
constants for various ratios of the overlap (of the Wannier functions)
and screening parameters: in general $U>V>X>W$, but in the extreme
screening limit $U>|X|>V=W$. In most real problems, the effect of the
exchange and pair hopping ($W$) can be neglected and we shall assume
this fact as valid also for an alternating chain. The phase diagram
corresponding to a half-filled upper band (see Fig. 3) in this case
($W=0$) and in terms of the original coupling constants
(with $A/B=3/4$, as for the $CuO_{3}$ chain) looks like in Fig. 5.
The oblique heavy line from Fig. 5 is defined by $Y_1=Y_2$ [see Eq. (32)]
(the dashed line corresponds to $Y_1=-Y_2$) and it is independent of the
inter-site coupling constant $V$; its slope depends only on the ratio $A/B$.
The two heavy halflines (horizontal and vertical) start from the point
Q, sliding on the dashed line by changing $V$ (upwards by decreasing $V$;
at $V=0$, $Q$ coincides with $P$); they delimit the region where a
superconductor state can occur. As can be remarked from Fig. 5 (drawn in
the particular case $X_A=X_B=-V=0.2$ eV) the region of $SS$ and $TS$
(denoted by I in Fig. 3) disappears for $V>0$. The values of the inter-site
and bond-site coupling constants for which the 3/4-filled undistorted
$CuO_{3}$ chain is superconductor can be determined also from Fig. 5;
for example, in the particular case of $V=0$, a superconductor state can
be reached only if $X_{A}/U_{A}>B/(4\sqrt{2}A)\simeq 0.24$ and
$X_{B}/U_{B}>A/(4\sqrt{2}B)\simeq 0.13$ [See Eq. (32)].
For a non-alternating chain, the same condition requires
$Y<0$, i.e. $X_{0}/U_{0}>1/(4\sqrt{2})\simeq 0.18$~\cite{B}
[See Eq. (33), where we replaced the renormalized coupling constants
${\cal U}$ and ${\cal X}$ by respectively $U_{0}$ and $X_{0}$ of a
non-alternating chain].
In order to get a superconductor state, we thus need a bigger ratio of
the ($A$) bond- and on-site couplings in the alternating case than in the
non-dimerized one; however, in average, the ratio is practically the same
and we could conclude that, within the present formalism,
the alternating structure does not play an
essential role in the occurrence of superconductivity in this system.
Let us also note that for a non-alternating structure and in the extreme
screening limit (when $V$ can be neglected), the ratio
(computed from atomic orbitals) $X_{0}/U_{0}$ has been estimated in the
range 0.15 -- 0.18 ~\cite{CGL}; and it is closed to the value required for a
superconductor state.

Finally, let us note that for realistic values of the parameters,
there is a coexistence of $CDW$ and $SDW$ fluctuations in the system
(see the shaded rectangle from Fig. 5). This phase could be interpreted
as follows: any allowed density modulation has a wave vector $2k_{F}$ and
consequently for the half-filled upper band the length wave of the density
fluctuation is equal to two lattice constants. From the alternating
structure it follows we could imagine $SDW$ and $CDW$ as located on
$A$ and $B$ sublattices respectively (a spin density wave requires
a stronger on-site repulsion than a charge density one).

%%%%%%%%%%%%%%%%%%%%%%%%%%%%%%%%%%%%%%%%%%%%%%%%%%%%%%%%%%%%%%%%%%%%%%%%%%%%%

\section{Conclusions}

The results of this paper can be summarized
as follows: the canonical transformation diagonalizing the one-particle
Hamiltonian for an alternating chain with two non-equivalent sites
per unit cell has been used
to find the expressions of the Hubbard-type interactions, initially
introduced in the site-representation, in the corresponding two-band model;
the consideration has been restricted to the particular case of only
alternating on-site energies (and equal hopping amplitudes) and a
gap between the two bands large enough to take into account
only the processes inside the partial filled band
(and we chose the upper band, as for the 3/4-filled $CuO_{3}$ chain).
A particularity of the alternating structure manifests itself in the
dependence of the obtained potentials on the momentum conservation
in a given process, i.e. if it is normal or umklapp one.
The RG method (second order) has then been applied to this (one-band)
model Hamiltonian; all the $g$-constants have been obtained by evaluating
the corresponding Hubbard-type interactions around the Fermi points.
The resulting ground-state phase diagrams have been analyzed
in terms of all coupling constants and for an arbitrary band filling.
The effect of the bond-site interaction is to renormalize the on-site
one. In the chosen coordinates (intra-bond, inter-site) and at half filling,
a critical inter-site
interaction controls the position of the ``charge-line" (separating a
dominant superconductor region from a density fluctuation one), and a critical
intra-bond coupling determines the ``spin-line" (separating a region
of singlet states from one with possible triplet states); the effect of
the density is to rotate these two lines and to change the position of
their crossing determined by a certain relation between the on-site
and bond-site coupling constants.
The particular case of the 3/4-filled
undistorted $CuO_{3}$ chain has been discussed in a separate section,
where the corresponding phase diagram has been drawn in terms of the
original coupling constants (neglecting $W$ term, usually very small);
the values of the inter-site and bond-site couplings required to get
a superconductor state (according to our approximate description)
have been also determined.

\acknowledgments
The authors are grateful to G. Japaridze for helpful suggestions
and stimulating discussions on this subject.
One of the authors (G.J.) acknowledges the support from the Russian
Foundation for Fundamental Researches (Grant No. 96--02--17527) and
from the INTAS--RFBR Program (Grant No. 95--591).

%\twocolumn[\hsize\textwidth\columnwidth\hsize\csname %
%@twocolumnfalse\endcsname

\begin{table}\widetext
\caption{The values of the $g$-constants for the upper band of a
chain with alternating on-site energies and Hubbard-type interactions
(last column), compared with the similar quantities for a non-alternating
structure (middle column). ${\cal U}$, ${\cal V}$, ${\cal X}$ and ${\cal W}$
are defined in the text by Eq. (21) and they are analogous, for our model,
of the Hubbard-like coupling constants $U_{0}$, $V_{0}$, $X_{0}$ and
$W_{0}$ from the usual case. By $n$ we denoted the electron filling of the
band, related to the Fermi momentum by $ak_{F}=\pi (1-n)$ for the upper
band of the alternating chain, or by $ak_{F}=\pi n$ in the
non-alternating case ($0\leq n \leq 1$).}
\begin{tabular}{lcc}
 & non-alternating chain & alternating chain \\ \tableline
$g^{}_{1{\scriptscriptstyle \parallel}}$
   & $2V_{0}\cos(2\pi n)+2W_{0}$
       & $-2{\cal V}\cos(\pi n)+2{\cal W}$    \\
$g^{}_{1\perp}$
   & $U_{0}+2V_{0}\cos(2\pi n)+8X_{0}\cos(\pi n)+4W_{0}$
     &  ${\cal U}-2{\cal V}\cos(\pi n)-
       8{\cal X}\sin(\pi n/2)+4{\cal W}$ \\
$g^{}_{2{\scriptscriptstyle \parallel}}$
   & $2V_{0}+2W_{0}\cos(2\pi n)$
     &  $2{\cal V}-2{\cal W}\cos(\pi n)$ \\
$g^{}_{2\perp}$
   & $U_{0}+2V_{0}+8X_{0}\cos(\pi n)+2W_{0}\{1+ \cos(2\pi n)\}$
     &  ${\cal U}+2{\cal V}-
       8{\cal X}\sin(\pi n/2)+2{\cal W}\{1-\cos(\pi n)\}$ \\
$g^{}_{3}$
   & $( U_{0}-2V_{0}-4W_{0}) \delta_{n,1/2}$
     &  $\{(U_{A}B^{4}-4\sqrt{2}X_{A}AB^{3})
	- (A\leftrightarrow B)\} \delta_{n,1/2}$ \\
$g^{}_{4}$
   & $U_{0}+2V_{0}+8X_{0}\cos(\pi n)+2W_{0}\{1+ \cos(2\pi n)\}$
     & ${\cal U}+2{\cal V}-
       8{\cal X}\sin(\pi n/2)+2{\cal W}\{1-\cos(\pi n)\}$
\end{tabular}
\end{table}

%]

%\narrowtext

\begin{table}
\caption{The scaling of the considered model (14) to one of the
two exactly soluble models -- Tomonaga-Luttinger (TL) or Luther-Emery (LE),
in each of the four regions defined by Eq. (27) in both the charge (c)
and spin (s) sector.}
\begin{tabular}{ccccc}
 & \multicolumn{2}{l}{half filling ($n=1/2$)}
	     & \multicolumn{2}{l}{not half filling ($n\neq 1/2$)} \\
 & c & s & c & s \\   \tableline
I & TL & TL & TL & TL \\
II & LE & TL & TL & TL \\
III & LE & LE & TL & LE \\
IV & TL & LE & TL & LE
\end{tabular}
\end{table}

\begin{figure}
\caption{Alternating chain with two non-equivalent sites per unit cell.}
\end{figure}

\begin{figure}
\caption{The two bands corresponding to
the one-particle tight binding Hamiltonian for an
alternating chain with two non-equivalent sites per unit cell.}
\end{figure}

\begin{figure}
\caption{The phase diagram corresponding to a chain with alternating
on-site atomic energies and Hubbard-type interactions
for a half-filled upper band. ${\cal W}_{1}=-Y/4$ and
${\cal V}_{2}=-Max \left\{ Y_{1},Y_{2} \right\}/2$ [see Eqs. (32) and
(33) in the text].The response functions corresponding
to the phases shown in the parentheses have a lower degree of
divergence than the others.}
\end{figure}

\begin{figure}
\caption{The phase diagram of the considered model away from half filling.
${\cal W}_{1}=-Y/4$ and ${\cal V}_{1}=Y/[2\cos(\pi n)]$ determine
the ``spin-line", while the ``charge-line" is determined by
${\cal W}_{2}=Y/[4\cos(\pi n)]$ and ${\cal V}_{2}=-Y/[4+2\cos(\pi n)]$
(here $Y>0$ and the band filling $n>1/2$).}
\end{figure}

\begin{figure}
\caption{The phase diagram corresponding to a half-filled upper
band of an alternating chain in terms of the original couplings
(site representation) and for a negligible intra-bond term; $P$ and $Q$
have the coordinates
$\left[
4(2)^{1/2} X_{A} A/B, 4(2)^{1/2} X_{B} B/A
\right]$
and
$\left[
4(2)^{1/2} X_{A} A/B-4VA^{2}/B^{2}, 4(2)^{1/2}
X_{B}B/A-4V B^{2}/A^{2}
\right]$
respectively and $ \tan \phi = B^{4}/A^{4} $. The picture has been
drawn in the particular case $A/B$=3/4, as for the $CuO_{3}$ chain where
the on-site couplings are estimated in the shaded rectangle.}
\end{figure}

\end{document}